\global\def\draftcontrol{0}

   \def\versionno{kappa}
\catcode`\@=11
\expandafter\ifx\csname draftcontrol\endcsname\relax\global\def\draftcontrol{0}
\fi

{\count255=\time\divide\count255 by 60
\xdef\hourmin{\number\count255}
\multiply\count255 by-60\advance\count255 by\time
\xdef\hourmin{\hourmin:\ifnum\count255<10 0\fi\the\count255}}
\def\draftdate{\number\month/\number\day/\number\year\ \ \ \hourmin }

\newcommand\makepapertitle{\par
  \begingroup
    \renewcommand\thefootnote{\@fnsymbol\c@footnote}%
    \def\@makefnmark{\rlap{\@textsuperscript{\normalfont\@thefnmark}}}%
    \long\def\@makefntext##1{\parindent 1em\noindent
            \hb@xt@1.8em{%
                \hss\@textsuperscript{\normalfont\@thefnmark}}##1}%
     \newpage
     \global\@topnum\z@   % Prevents figures from going at top of page.
     \@makepapertitle
     \thispagestyle{empty}\@thanks
  \endgroup
  \setcounter{footnote}{0}%
  \global\let\thanks\relax
  \global\let\makepapertitle\relax
  \global\let\@makepapertitle\relax
  \global\let\@thanks\@empty
  \global\let\@author\@empty
  \global\let\@date\@empty
  \global\let\@title\@empty
  \global\let\title\relax
  \global\let\author\relax
  \global\let\date\relax
  \global\let\and\relax
  \def\version{\let\version\@version\@gobble}
}
\def\@makepapertitle{%
  \newpage
   \ifnum\draftcontrol=1 {}
   \version\versionno
   \vskip 3em%
   \else
   \hfill\hbox to 3cm {\parbox{4cm}{\@pubnum}\hss}%
   \vskip 3em%
   \fi
   \begin{center}%
   \let \footnote \thanks
     {\LARGE {\@title}}%
     \vskip 1.5em%
     {\normalsize%\large
       \lineskip .5em%
       \begin{tabular}[t]{c}%
         \@author
       \end{tabular}\par}%
     \vskip 1.5em%
     {\@bstract}%
     \end{center}%
     \vskip 1.5em
     \@date%
   \par
}

\gdef\@pubnum{}
\def\pubnum#1{%
  \gdef\@pubnum{#1}}

\gdef\@bstract{}
\def\Abstract#1{%
  \gdef\@bstract{%
   \parbox{\textwidth-0pc}{%
   \centerline{\bf Abstract}\penalty1000%
\kern.2cm%
\noindent%\abstractfont \baselineskip=12pt
\renewcommand\baselinestretch{1.0}%
{#1}}}
}

\def\ps@paper{\let\@mkboth\@gobbletwo%
     \ifnum\draftcontrol=1
    \def\@oddfoot{\hbox to \textwidth{\tiny \versionno \hfil\tiny\draftdate}%
    \hskip -\textwidth \hbox to \textwidth{\hfil\rm\thepage\hfil}}%
     \else\def\@oddfoot{\hbox to \textwidth{\hfil\rm\thepage\hfil}}
     \fi
     \let\@evenfoot\@oddfoot
}
\def\body{\clearpage
          \pagestyle{paper}
    }
\def\@version#1{\ifnum\draftcontrol=1
\typeout{}\typeout{#1}\typeout{}
\vskip3mm\centerline{\hbox{\fbox{\normalsize{\tt DRAFT -- #1 -- }
                   {\draftdate}}}}\vskip3mm
\fi}
\let\version\@version
\long\def\eqlabel#1{\ifnum\draftcontrol=1
                    \tag@false  % there are some problems with multline without this
                    \tag*{(\theequation) \hbox to -0.2cm{\hspace{0cm}\small{#1}\hss}}
                    \refstepcounter{equation}
                    \edef\@currentlabel{\theequation}
                    \ltx@label{#1}          % use old LaTeX \label instead of new definition
                    \else
                    \label{#1}
                    \fi
                    }
\let\st@bibitem\@bibitem
\let\st@lbibitem\@lbibitem
\ifnum\draftcontrol=1
  \def\@bibitem#1{%
    \st@bibitem{#1}\a@@label{#1}\ignorespaces}
  \def\@lbibitem[#1]#2{%
    \st@lbibitem[#1]{#2}\a@@label{#2}\ignorespaces}
  \def\a@@label#1{%
    \gdef\a@lab{\smash{\normalfont\small#1}}
    \ifvmode
      \if@inlabel
        \global\setbox\@labels\hbox{%
          \llap{\a@lab\let\a@lab\relax
                \kern\@totalleftmargin\kern\marginparsep}%
          \box\@labels}%
      \fi
    \fi}
\fi
\documentclass[12pt,letterpaper]{article}
\usepackage{amsmath,amssymb,array,calc,epsfig,rotating,bm,xcolor}
\usepackage[sort]{cite}
\usepackage{graphicx,esint,float}
\usepackage{psfrag,verbatim}
\usepackage[makeroom]{cancel}
\usepackage{xcolor,url}
\usepackage{hyperref}
\ifnum\draftcontrol=1
\fi
\renewcommand\baselinestretch{1.25}
\renewcommand\section{\@startsection {section}{1}{\z@}%
                                   {-3.5ex \@plus -1ex \@minus -.2ex}%
                                   {2.3ex \@plus.2ex}%
                                   {\normalfont\large\bfseries}}
\renewcommand\subsection{\@startsection{subsection}{2}{\z@}%
                                   {-3.25ex\@plus -1ex \@minus -.2ex}%
                                   {1.5ex \@plus .2ex}%
                                   {\normalfont\normalsize\bfseries}}
\renewcommand\subsubsection{\@startsection{subsubsection}{3}{\z@}%
                                   {-3.25ex\@plus -1ex \@minus -.2ex}%
                                   {1.5ex \@plus .2ex}%
                                   {\normalfont\normalsize\it}}
\renewcommand\paragraph{\@startsection{paragraph}{4}{\z@}%
                                   {-3.25ex\@plus -1ex \@minus -.2ex}%
                                   {1.5ex \@plus .2ex}%
                                   {\normalfont\normalsize\bf}}

\numberwithin{equation}{section}

\def\revise#1       {\raisebox{-0em}{\rule{3pt}{1em}}%
                     \marginpar{\raisebox{.5em}{\vrule width3pt\
                     \vrule width0pt height 0pt depth0.5em
                     \hbox to 0cm{\hspace{0cm}{%
                     \parbox[t]{4em}{\raggedright\footnotesize{#1}}}\hss}}}}

\newcommand\nxt[1]  {\\\fnxt#1}
\newcommand{\ie}{{\it i.e.,}\ }
\newcommand{\eg}{{\it e.g.,}\ }

\def\calc         {{\cal C}}

\def\calf         {{\cal F}}

\def\calm         {{\cal M}}
\def\caln         {{\cal N}}
\def\calo         {{\cal O}}

\def\calr         {{\cal R}}

\def\del          {\partial}

\def\sqr#1#2{{\vcenter{\vbox{\hrule height.#2pt
 \hbox{\vrule width.#2pt height#1pt \kern#1pt
 \vrule width.#2pt}\hrule height.#2pt}}}}

\def\w{\omega}

\def\hg{\hat{\Gamma}}

\def\aa1{\phi}
\def\cc1{\psi}

\def\csb{{\chi\rm{SB}}}

\def\f0{\text{\boldmath$\varphi$}}
\def\h2{\mathfrak{h}}

\catcode`\@=12

\begin{document}

%%%
%%%%%% text starts here
%%%%%%%%%

\title{\bf Gravitational susceptibility of QGP}

\date{November 24, 2022}
%\date\today

\author{
Alex Buchel\\[0.4cm]
\it $ $Department of Physics and Astronomy\\ 
\it University of Western Ontario\\
\it London, Ontario N6A 5B7, Canada\\
\it $ $Perimeter Institute for Theoretical Physics\\
\it Waterloo, Ontario N2J 2W9, Canada
}

\Abstract{We use ${\cal N}=2^*$ and cascading gauge theory holographic models to
extract the general features of the gravitational susceptibility
$\kappa$ of strongly coupled nonconformal quark-gluon plasma. We show
that in theories with a relevant coupling constant the gravitational
susceptibility is renormalization scheme dependent. We propose to use
its temperature derivative, \ie $\frac{d\kappa}{d\ln T}$, as a
scheme-independent characteristic of a QGP. Although $\kappa$ is a
thermodynamic quantity, its critical behavior can be drastically
distinct in the vicinity of seemingly identical thermal phase
transitions.
}

\makepapertitle

\body

\version\versionno
\tableofcontents

\section{Introduction and summary}\label{intro}

Modern relativistic hydrodynamics \cite{Kovtun:2012rj} is a widely
accepted framework to analyze strongly coupled quark-gluon plasma (QGP) 
produced in high energy heavy-ion collisions
\cite{Kolb:2003dz,STAR:2005gfr,PHENIX:2004vcz,PHOBOS:2004zne,BRAHMS:2004adc}. 
It is an effective theory of the conservation law of the
fluid stress-energy tensor\footnote{We consider uncharged fluids here. The general
hydrodynamic treatment must include the conservation of all  conserved four-currents
$J^\mu_i$ of the theory.}
$T^{\mu\nu}$,
\begin{equation}
\nabla_\mu T^{\mu\nu}=0\,,\qquad T^{\mu\nu}=\sum_{n=0}^\infty T^{\mu\nu}_{(n)}\,,
\eqlabel{cons}
\end{equation}
organized as an expansion in the
gradients of its four-velocity $u^\mu$, $T^{\mu\nu}_{(n)}\sim \nabla^n u$.
It is deemed to be applicable close to equilibrium and in weakly curved
background space-times\footnote{As in most effective theories, the series expansion
in \eqref{cons} is asymptotic and has zero radius of convergence \cite{Heller:2013fn,Buchel:2016cbj}.}.
Specifically\footnote{We are using the
Landau-Lifshitz frame.}, at zero order in the gradients, the stress-energy tensor is that of the
thermal equilibrium of the theory in Minkowski space-time,
\begin{equation}
\begin{split}
&T^{\mu\nu}_{(0)}=(\epsilon+P)\ u^\mu u^\nu +P\ g^{\mu\nu}\ =\ \epsilon\ u^\mu u^\nu +P\ \Delta^{\mu\nu}\,,\\
&\Delta^{\mu\nu}\equiv g^{\mu\nu}+u^\mu u^\nu\,,\qquad g_{\mu\nu}u^\mu u^\nu =-1\,,
\end{split}
\eqlabel{t0}
\end{equation}
where the pressure $P$ is related to the energy density $\epsilon$ via the equilibrium
equation of state $P=P_{eq}(\epsilon)$, and $g^{\mu\nu}$ is the background space-time
metric tensor. Additionally, the local temperature $T$ and the entropy density
$s$ are introduced as
\begin{equation}
\epsilon +P= sT\,,\qquad d\epsilon=T ds\,.
\eqlabel{stlocal}
\end{equation}
At the first-order in the velocity gradients
constitutive relations between the stress-energy tensor $T^{\mu\nu}_{(1)}$
and the four-velocity require
two-independent transport coefficients --- the shear $\eta$, and the bulk $\zeta$ viscosities:
\begin{equation}
\begin{split}
&T^{\mu\nu}_{(1)}=-\eta\ \sigma^{\mu\nu}-\zeta\ \Delta^{\mu\nu}\nabla_\alpha u^\alpha\,,\\
&\sigma^{\mu\nu}\equiv \Delta^{\mu\alpha}\Delta^{\nu\beta}\biggl(\
\nabla_\alpha u_\beta+\nabla_\beta u_\alpha-\frac 23 \Delta_{\alpha\beta}\nabla_\gamma u^\gamma
\ \biggr)\,.
\end{split}
\eqlabel{t1}
\end{equation}
The sensitivity of the fluid to the space-time background curvature arises
at the second-order in the velocity gradients. At the second-order in the
gradient expansion, there are 5
second-order transport coefficients, if the fluid is conformal \cite{Baier:2007ix},
and 15  coefficients for a general nonconformal theory \cite{Romatschke:2009kr}. 
In this paper we will be interested in the  gravitational coupling
of the general hydrodynamics, so we present only the relevant terms of the
Romatschke classification \cite{Romatschke:2009kr}:
\begin{equation}
\begin{split}
T^{\mu\nu}_{(2),grav}=&\kappa\ \biggl(\
R^{\langle\mu\nu\rangle}-2u_\alpha u_\beta\ R^{\alpha\langle\mu\nu\rangle\beta}
\ \biggr) +2 \kappa^*\ u_\alpha
u_\beta\ R^{\alpha\langle\mu\nu\rangle\beta}\\
&+\Delta^{\mu\nu}\biggl(\
\zeta_5\ R+\zeta_6\ u^\alpha u^\beta\ R_{\alpha\beta}
\
\biggr)\,,
\end{split}
\eqlabel{t2g}
\end{equation}
where
\begin{equation}
R^{\mu\langle\nu\alpha\rangle\beta}\equiv \frac 12 R^{\mu\kappa\sigma\beta}
\biggl(\ \Delta_\kappa^\nu\Delta_\sigma^\alpha+\Delta_\sigma^\nu\Delta_\kappa^\alpha-\frac 23
\Delta^{\nu\alpha}\Delta_{\kappa\sigma}
\
\biggr)
\eqlabel{rbrac}
\end{equation}
is constructed from the curvature tensor $R^{\mu\nu\alpha\beta}$ of the background
metric $g^{\mu\nu}$. Of the four gravitational transport coefficients
$\{\kappa$, $\kappa^*$, $\zeta_5$, $\zeta_6\}$ only the gravitational
susceptibility $\kappa$ is independent: non-negativity
of the entropy current divergence requires
\cite{Romatschke:2009im,Bhattacharyya:2012nq,Jensen:2012jh}
\begin{equation}
\begin{split}
&\kappa^*=\kappa-\frac T2\ \frac{d\kappa}{dT}\,,\\
&\zeta_5=\frac12 \biggl(\
c_s^2 T\ \frac{d\kappa}{dT}-c_s^2\kappa-\frac \kappa3
\
\biggr)\,,\\
&\zeta_6=c_s^2\biggl(\
3T\ \frac{d\kappa}{dT}-2T\ \frac{d\kappa^*}{dT}+2\kappa^*-3\kappa
\ \biggr)-\kappa+\frac43 \kappa^*+\frac{\lambda_4}{c_s^2}\,,
\end{split}
\eqlabel{3rel}
\end{equation}
where $c_s$ is the speed of the sound wave
\begin{equation}
c_s^2=\frac{dP}{d\epsilon}\,,
\eqlabel{defcs}
\end{equation}
and $\lambda_4$ is the second-order nonlinear transport coefficient appearing in $T^{\mu\nu}_{(2)}$
as \cite{Romatschke:2009kr}
\begin{equation}
T^{\mu\nu}_{(2)}=\cdots +\lambda_4\ \nabla^{\langle\mu}\ln s\nabla^{\nu\rangle}\ln s+\cdots\,.
\eqlabel{l4app}
\end{equation}
Both $\kappa$ and $\lambda_4$ are thermodynamic quantities and can be extracted from
the Euclidean (correspondingly) 2- and 3-point correlation functions of the stress-energy
tensor\footnote{See also \cite{Kovtun:2018dvd,Shukla:2019shf}.} \cite{Moore:2012tc}
\begin{equation}
\begin{split}
&\kappa=\lim_{k_z\to 0}\ \frac{\del^2}{\del k_z^2}G_E^{xy,xy}(k)\bigg|_{k_0=0}\,,\\
&\lambda_4=-2\kappa^*+\kappa-\frac{c_s^4}{2}\lim_{p^x,q^y\to 0}\frac{\del^2}{\del p_x\del q_y}G_E^{tt,tt,xy}(p,q)\bigg|_{p_0,q_0=0}\,.
\end{split}
\eqlabel{corr}
\end{equation}

In this paper we will be interested in the gravitational susceptibility $\kappa$
of a QGP. 
Being a thermodynamic coefficient, it can, in principle, be computed from the
corresponding gauge theory lattice implementation \cite{Moore:2012tc}. Instead, we use
the holographic correspondence \cite{Maldacena:1997re,Aharony:1999ti}
and extract $\kappa$ from the retarded correlation function of the gauge theory
stress-energy tensor \cite{Baier:2007ix}. 
We focus on two examples of holographic models:
\begin{itemize}
\item  the mass-deformed $\caln=4$ supersymmetric $SU(N)$ Yang-Mills theory
also know as $\caln=2^*$ gauge theory \cite{Pilch:2000ue,Buchel:2000cn};
\item  the $\caln=1$ supersymmetric $SU(N+M)\times SU(N)$ cascading
gauge theory \cite{Klebanov:2000hb}.
\end{itemize}
Both theories are nonconformal --- in the former, the scale invariance
is broken explicitly by the mass terms for the bosonic and the fermionic
components of the $\caln=2$ hypermultiplet; in the latter,
the scale invariance is broken spontaneously through the dimensional transmutation
of the gauge couplings. 
Our holographic models
are examples of top-down\footnote{While some observables, \eg the ratio
of the shear viscosity to the entropy density, are universal in all
holographic models in the supergravity approximation \cite{Buchel:2003tz},
certain exotic phase transitions are ubiquitous in phenomenological holography,
but not in string theory \cite{Buchel:2020xdk,Buchel:2022zxl}.},
rather than phenomenological, holography.

Before we report our result, we review what is known in the literature. 
\nxt The gravitational susceptibility of $\caln=4$ $SU(N)$ SYM in the planar
limit and at infinitely large 't Hooft coupling constant  was computed
in \cite{Baier:2007ix}
\begin{equation}
\kappa\bigg|_{\caln=4} = \frac{T^2N^2}{8}\qquad \Longrightarrow\qquad 
\frac{4\pi^2 \kappa T}{s}\bigg|_{\caln=4}=1\,,\qquad \frac{2\pi^2 T^2}{s}
\frac{d\kappa}{dT}\bigg|_{\caln=4}=1\,,
\eqlabel{ksym}
\end{equation}
where we also presented two benchmark quantities that would allow for comparison
with other models. 
\nxt The finite 't Hooft coupling corrections for the $\caln=4$ QGP $\kappa$ were evaluated
in \cite{Buchel:2008bz}
\begin{equation}
\frac{4\pi^2 \kappa T}{s}\bigg|_{\caln=4}=1-\frac{265}{8}\zeta(3)\ (g_{YM}^2 N)^{-3/2}+\cdots\,.
\eqlabel{ksymf}
\end{equation}
\nxt For weakly coupled $SU(N)$ gauge theory \cite{Romatschke:2009ng}
\begin{equation}
\frac{4\pi^2 \kappa T}{s}\bigg|_{SU(N),{\rm free}}=\frac 52\,.
\eqlabel{weakly}
\end{equation}
\nxt $\kappa$ was determined directly to the leading order in lattice perturbation
theory for QCD QGP in \cite{Philipsen:2013nea}.
\nxt Using large-$N$ QFT techniques, the computation of $\kappa$ were performed for the
$\rm{O(N)}$ model for any coupling value in \cite{Romatschke:2019gck}.
\nxt $\kappa$ was computed in certain phenomenological nonconformal models in
\cite{Finazzo:2014cna}. However, the validity of the results presented there should be
verified with the implementation of the holographic renormalization ---
at least for the class of models we consider here the proper treatment of the holographic
renormalization, including the finite counterterms and the corresponding issue of the
scheme dependence, is crucial to obtain correct results.

We now summarize our results:
\begin{itemize}
\item It is well known that in a quantum field theory (QFT) the expectation value of the
stress-energy tensor is renormalization scheme dependent. From the holographic
perspective, renormalization of the boundary correlation functions is sensitive
to finite counterterms --- one can crudely think that the energy density and the pressure of 
$T_{(0)}^{\mu\nu}={\rm diag}\{\epsilon,P,P,P\}$ in \eqref{t0} are defined up to  additive 
constants. If the theory, as well as its regularization and the renormalization, preserves
supersymmetry, some of the finite counterterms can be fixed requiring the vanishing
of the stress-energy tensor expectation value in a supersymmetric vacuum state, \ie as
$T\to 0$, see \cite{Buchel:2004hw} for example. For a QFT in curved space-time and/or with time-dependent relevant couplings,
there are even more possibilities for finite counterterms and thus the renormalization
scheme dependence\footnote{See  a discussion of this issue in the holographic context in
\cite{Buchel:2012gw}.}. These new counterterms can not be fixed requiring Minkowski space-time
supersymmetry. As we explicitly show in section \ref{framework}, holographic models
with relevant couplings for dimension $\Delta=\{2,3\}$ operators --- \eg the mass terms
$\{m_b^2,m_f\}$
for the bosons and fermions --- introduce the scheme dependence for $\kappa$. Specifically,
the gravitational susceptibility in such models is defined up to arbitrary
constants $\{\delta_f,\delta_b\}$:
\begin{equation}
\kappa\qquad \longrightarrow\qquad \kappa+\delta_f\ m_f^2++\delta_b\ m_b^2\,.
\eqlabel{schemek}
\end{equation}
The corresponding finite counterterms involve the boundary curvature tensor, and thus are insensitive
to Minkowski supersymmetry.
From \eqref{schemek} it is clear that the renormalization scheme independent
quantity is $\frac{d\kappa}{d\ln(T)}$; to this end we propose to characterize the gravitational
susceptibility of QGPs with $\calr_\kappa$,
\begin{equation}
\calr_\kappa\equiv 2\pi^2\ \frac{T}{s}\ \frac{d\kappa}{d\ln T}= 4\pi^2\ \frac{T}{s}\ (\kappa-\kappa^*)\,, 
\eqlabel{defcalrk}
\end{equation}
where the normalization is chosen with \eqref{ksym} in mind.
\item Supersymmetry tames somewhat the value of $\calr_\kappa$ \eqref{defcalrk}:
in supersymmetric $\caln=2^*$ theory with $m_b^2=m_f^2$, 
\begin{equation}
\calr_\kappa\bigg|_{\caln=2^*,m_b=m_f}\ \in\ \left[1,\frac 54\right]\,,\qquad {\rm as}\qquad
\frac{m_b^2}{T^2}\in [0,+\infty) \,,
\eqlabel{rrangen2}
\end{equation}
and in $\caln=1$ supersymmetric cascading gauge theory  $\calr_\kappa$ grows as
\begin{equation}
\calr_\kappa\bigg|_{cascading}\ \in\ \biggl[1,2.23(1)\biggr]\,,\qquad {\rm as}\qquad
{T}\in \left[{T_\csb},+\infty\right)\,,\qquad  T_\csb=0.541(9)\Lambda\,,
\eqlabel{rrangekt}
\end{equation}
where $\Lambda$ is the strong coupling scale of the cascading gauge theory.
The results reported are for the cascading QGP with the unbroken chiral symmetry ---
this phase becomes perturbatively unstable to chiral symmetry breaking fluctuations
below $T_\csb$ \cite{Buchel:2010wp}. We can not use holography to compute
$\kappa$ in the confining phase of the theory, which occurs, as a large-$N$ suppressed
first-order phase transition, for $T<T_{c}=0.614(1)\Lambda$ \cite{Aharony:2007vg}.
The deconfined phase of the cascading gauge theory with spontaneously broken
chiral symmetry is unstable to energy density fluctuations (the sound waves)
\cite{Buchel:2018bzp}, thus we
do not report the susceptibility in this phase as well. 
\item $\caln=2^*$ gauge theory with $m_b\ne 0$ and $m_f=0$ completely breaks the supersymmetry.
Here we find
\begin{equation}
\calr_\kappa\bigg|_{\caln=2^*,m_f=0}\ \in\ \left[1,-\infty\right)\,,\qquad {\rm as}\qquad
\frac{m_b^2}{T^2}\in [0,5.4(1)] \,.
\eqlabel{rrangen2b}
\end{equation}
The thermal deconfined states of the theory exist only for $T>T_{crit}=2.3(3) m_b$.
In the vicinity of the critical point, \ie as $T\to T_{crit}+0$, the speed of the sound waves vanishes
and the specific heat diverges \cite{Buchel:2007vy} 
\begin{equation}
c_s^2\propto\ \pm\ \sqrt{T-T_{crit}}\,,\qquad c_{\rm V}\propto\ \pm\ \frac{1}{\sqrt{T-T_{crit}}}\,,
\eqlabel{n2crit}
\end{equation}
and we further find, see section \ref{n2qgp},
\begin{equation}
\calr_\kappa\bigg|_{\caln=2^*,m_f=0}\ \propto\  \mp\ \frac{1}{\sqrt{T-T_{crit}}}\,.
\eqlabel{rcrit}
\end{equation}
The signs in \eqref{n2crit} and \eqref{rcrit} correlate: there is an additional deconfined phase
of the theory (the lower signs), which is however unstable to sound waves, $c_s^2<0$.
\item Cascading gauge theory plasma in the chirally symmetric phase has an identical
critical point to that of $\caln=2^*$ \eqref{n2crit}. The analogous (terminal) temperature here is
$T_u=0.537(3)\Lambda$ \cite{Buchel:2009bh}, and as $T\to T_u+0$ we have
\begin{equation}
c_s^2\propto\ \pm\ \sqrt{T-T_{u}}\,,\qquad c_{\rm V}\propto\ \pm\ \frac{1}{\sqrt{T-T_{u}}}\,.
\eqlabel{casu}
\end{equation}
Once again, there are two deconfined phases, both existing only for $T>T_u$, that join
at the terminal temperature $T_u$. Interestingly, we find that despite identical critical thermodynamics
of $\caln=2^*$ and the cascading QGP,
the gravitational susceptibility of the cascading gauge theory plasma at criticality is very different
(see section \ref{cqgp})
\begin{equation}
\calr_\kappa\bigg|_{cascading}=\ \underbrace{{\rm const}}_{>0}\  \mp\ \propto\ \sqrt{T-T_{u}}\,.
\eqlabel{ru}
\end{equation}
As we explore in more details in section \ref{cqgp}, as $T\to T_u+0$,
\begin{equation}
\frac{\kappa}{T^2}\bigg|_{cascading}=\calc_0+\calc_1\ (T-T_u)\pm\calc_2\ (T-T_u)^{3/2}+\cdots\,,
\eqlabel{ktkks}
\end{equation}
where  $\calc_2>0$. 
Note that $\kappa$ of the cascading gauge theory plasma is scheme independent as the theory
lacks relevant couplings.
 \end{itemize}

Within our computational framework, see section \ref{framework}, we will also have access to
the {\it shear} relaxation time $\tau_\pi$. 
This  second-order transport coefficient  enters $T^{\mu\nu}_{(2)}$ as \cite{Baier:2007ix}
\begin{equation}
T^{\mu\nu}_{(2)}=\cdots+\eta\tau_\pi \left(\
u\cdot \nabla\sigma^{\mu\nu}+\frac{\nabla\cdot u}{3}\sigma^{\mu\nu}
\ \right)
+\cdots\,.
\eqlabel{t2tau}
\end{equation}
\nxt The shear relaxation time was computed for the $\caln=4$ $SU(N)$ SYM plasma in the
planar limit and at infinitely large 't Hooft coupling constant in \cite{Baier:2007ix}
\begin{equation}
T\tau_\pi\bigg|_{\caln=4}=\frac{2-\ln 2}{2\pi}\,.
\eqlabel{taupin4}
\end{equation}
\nxt The finite 't Hooft coupling corrections for the $\caln=4$ QGP $\tau_\pi$
were evaluated in \cite{Buchel:2008bz}
\begin{equation}
T\tau_\pi\bigg|_{\caln=4}=\frac{2-\ln 2}{2\pi}+\frac{375}{32\pi}\zeta(3)\ (g_{YM}^2N)^{-3/2}+\cdots\,.
\eqlabel{taupin4cor}
\end{equation}

To report results obtained in this work we introduce
\begin{equation}
\calr_{\tau_\pi}\equiv \frac{2\pi}{2-\ln 2}\ T\tau_\pi\,,
\eqlabel{defrtau}
\end{equation}
where the normalization is chosen with \eqref{taupin4} in mind. 
\begin{itemize}
\item Unlike the gravitation susceptibility, the shear relaxation time of a
QGP is free from renormalization scheme ambiguities, see section \ref{framework}
for details.
\item In the supersymmetric $\caln=2^*$ theory with $m_b^2=m_f^2$,
\begin{equation}
\calr_{\tau_\pi}\bigg|_{\caln=2^*,m_b=m_f}\in \biggl[1,1.1(5)\biggr]\,,\qquad {\rm as}\qquad
\frac{m_b^2}{T^2}\in [0,+\infty)\,,
\eqlabel{rtaun2susy}
\end{equation}
while in the cascading gauge theory
\begin{equation}
\calr_{\tau_\pi}\bigg|_{cascading}\in \biggl[1,1.8(8)\biggr]\,,\qquad {\rm as}\qquad
T\in[T_\csb,+\infty)\,,
\eqlabel{rtaucas}
\end{equation}
where $T_\csb$ is given in \eqref{rrangekt}.
\item In the vicinity of the critical point \eqref{n2crit}, the relaxation time of the $\caln=2^*$
plasma diverges as (see section \ref{n2qgp})
\begin{equation}
\calr_{\tau_\pi}\bigg|_{\caln=2^*,m_f=0}\propto \mp \frac{1}{\sqrt{T-T_{crit}}}\,.
\eqlabel{taucritn2}
\end{equation}
Note that the minus sign occurs on the thermodynamic branch with $c_s^2>0$; we find that
\begin{equation}
T\tau_\pi\bigg|_{\caln=2^*,m_f=0}<0\qquad {\rm for}\qquad \frac{T}{T_{crit}}\in (\ 1,1.003(0)\ )\,.
\eqlabel{taucritn21}
\end{equation}
Given that the model discussed is a top-down holography, it would be extremely interesting
to study whether a negative shear relaxation time implies some
instabilities and/or causality violations. Typically, a combination of transport coefficients
appears in physical observables. For example, the dispersion relation of the sound waves takes the
form
\begin{equation}
\omega=\pm c_s q-i \Gamma q^2\pm \frac{\Gamma}{c_s}\left(c_s^2\tau_{eff} -\frac \Gamma2\right)q^3+\calo(q^4)\,,
\eqlabel{dosps}
\end{equation}
where
\begin{equation}
\Gamma=\frac{2\eta}{3 sT}+\frac{\zeta}{2sT}\,,\qquad \tau_{eff}=\frac{\tau_\pi
+\frac 34\frac\zeta\eta \tau_\Pi}{1+\frac 34\frac\zeta\eta}\,,
\eqlabel{tg}
\end{equation}
with $\tau_\Pi$ being the {\it bulk} relaxation time \cite{Romatschke:2009kr}. It was determined
in \cite{Buchel:2009hv} that $\tau_{eff}>0$ in $\caln=2^*$ QGP in the phase with $c_s^2>0$,
and diverges as $\tau_{eff}T\propto \pm (1-T_{crit}/T)^{-1/2}$ in the critical region \eqref{n2crit}.
\item The shear relaxation time of the cascading QGP is positive and finite, but is not analytic in the
critical region \eqref{casu} (see section \ref{cqgp})
\begin{equation}
\calr_{\tau_\pi}\bigg|_{cascading}\ =\ \underbrace{\rm const}_{>0}\ \mp\ \sqrt{T-T_u} \,.
\eqlabel{castaucr}
\end{equation}
\end{itemize}

The rest of the paper is organized as follows. In section \ref{framework} we review
the holographic framework used to compute $\kappa$. We explain why $\kappa$,
but not $\tau_\pi$, is renormalization scheme dependent in theories with
$\Delta=\{2,3\}$ relevant couplings. 
We discuss $\kappa$ and $\tau_\pi$ of the cascading gauge theory and $\caln=2^*$ QGPs in sections
\ref{cqgp} and \ref{n2qgp} correspondingly.

\section{Holographic computation of $\kappa$}
\label{framework}

Given the second-order formulation of the relativistic hydrodynamics reviewed in section \ref{intro},
the $(xy,xy)$-component of the retarded stress-energy tensor Green's function
in the limit of the small frequency $\omega$ and the small momentum $q=|\vec{q}|$ takes the form
\cite{Romatschke:2009kr}
\begin{equation}
G_R^{xy,xy}(\omega,q)=P-i\eta\ \omega
+\underbrace{\biggl(\eta\tau_\pi-\frac\kappa2+\kappa^*\biggr)}_{\equiv T^2\hg_\omega}\ \omega^2
+\underbrace{\biggl(-\frac\kappa2\biggr)}_{\equiv T^2\hg_q}\ q^2
+\calo(\omega q^2,\omega^3)\,,
\eqlabel{gr}
\end{equation}
where we introduced dimensionless quantities $\hg_\omega$ and $\hg_{q}$.
Notice that
\begin{equation}
T^2 \left(\hg_{\omega}+\hg_{q}\right)=\eta\tau_\pi +\kappa^*-\kappa=\eta\tau_\pi -
\frac T2\frac{d\kappa}{dT} \,,
\eqlabel{gsum}
\end{equation}
where we used \eqref{3rel}.

The computation of the Green's function \eqref{gr} in holography was explained in
\cite{Son:2002sd}:
\begin{itemize}
\item Consider the five-dimensional bulk gravitational action $S_{bulk}$, dual to some boundary QFT.
The thermal equilibrium state of the boundary gauge theory is dual to a black brane geometry,
\begin{equation}
ds_5^2=-c_1^2\ dt^2+c_2^2\ {\bm x}^2 + c_3^2\ d\rho^2\,,
\eqlabel{5dmetric}
\end{equation}
where $c_i=c_i(\rho)$ are functions of the radial coordinate $\rho$. We assume that $\rho\to 0$ is the
asymptotic boundary, while $\rho\to \rho_H$ is a regular Schwarzschild horizon,
\begin{equation}
\lim_{\rho\to 0} \frac{c_1}{c_2}=1\,,\qquad \lim_{\rho\to \rho_H} c_1 = 0\,.
\eqlabel{metlim}
\end{equation}
\item The retarded correlation function $G_R^{xy,xy}$ can be extracted from the quadratic boundary
effective action for the metric fluctuations $\varphi(t,z,\rho)\equiv \frac 12 c_2^{-2} \delta g_{xy}(t,z,\rho)$,
\begin{equation}
\varphi^b(\omega,q)=\int {d\omega dq}\ e^{i \omega t - i q z }\varphi(t,z,\rho)\bigg|_{\rho\to 0}\,,
\eqlabel{pres1}
\end{equation}
given by
\begin{equation}
S_{boundary}[\varphi^b]=\int \frac{d\omega dq}{(2\pi)^2}\ \varphi^b(-\omega,-q)\ \calf(\omega,q)\ \varphi^b(\omega,q)\,,
\eqlabel{pres2}
\end{equation}
as
\begin{equation}
G_R^{xy,xy}(\omega,q)= -2\ \calf(\omega,q)\,.
\eqlabel{pres3}
\end{equation}
\item The boundary metric functional in \eqref{pres2} is defined as
\begin{equation}
S_{boundary}[\varphi^b]=\lim_{\rho\to 0}\biggl(\
S_{bulk}^\rho[\varphi]\ +\ S_{GH}[\varphi]\ +\ S^{counter}[\varphi]
\ \biggr)\,,
\eqlabel{pres4}
\end{equation}
where $S_{bulk}^\rho$ is the regularized bulk gravitational action, evaluated
on-shell for the bulk metric fluctuation $\varphi$, subject to the following boundary
conditions:
\begin{equation}
\begin{split}
&(a):\qquad \lim_{\rho\to 0} \varphi(t,z,\rho)=\varphi^b(t,z)\,;\\
&(b):\qquad \varphi(t,z,\rho)\ {\rm is\ an\ incoming\ wave\ at\ the\ horizon,\ \ie as}\
\rho\to\rho_H\,.
\end{split}
\eqlabel{pres5}
\end{equation}
Also, $S_{GH}$ is the standard Gibbons-Hawking term over the regularized boundary.
The purpose of the boundary counterterm $S^{counter}$ is to remove divergences
of the regularized boundary action $S_{bulk}^\rho+S_{GH}$ as 
$\rho\to 0$, rendering the renormalized boundary action \eqref{pres4} finite.
\end{itemize}

To evaluate $\hg_\omega$ and $\hg_q$ we need the boundary functional \eqref{pres4}
to quadratic order in $\calo(\w^2,q^2)$ --- thus we need the on-shell solution for
$\varphi$ to this order as well. It was shown in \cite{Buchel:2004qq} that the equation
for $\varphi$ is simply that of the minimally coupled massless scalar in the background
metric \eqref{5dmetric}. Furthermore, the solution can be expanded as \cite{Buchel:2004qq}
\begin{equation}
\phi(t,z,\rho)=e^{-i\omega t+i q z}\ \underbrace{\phi^b(\omega,q)}_{(a)}\
\underbrace{\left(\frac{c_1}{c_2}\right)^{-i \omega Q}}_{(b)}\ \biggl(\
1+\omega^2\     z_2(\rho)+q^2\ z_3(\rho)+\calo(\omega q^2,\omega^3) 
\ \biggr)\,,
\eqlabel{hydro}
\end{equation}
where
\begin{equation}
Q\equiv \frac{1}{2\pi T}\,.
\eqlabel{defq}
\end{equation}
with $T$ begin the Hawking temperature of the black brane \eqref{5dmetric}.
In \eqref{hydro} we highlighted components of the solution that take care of the boundary conditions
\eqref{pres5}. The radial functions $\{z_2$, $z_3\}$ satisfy 
\begin{equation}
\begin{split}
&0=z_2''+\left(\ln\frac{c_1c_2^3}{c_3}\right)'\ z_2'+\frac{c_3^2}{c_1^2}-Q^2\
\left[\left(\ln \frac{c_1}{c_2}\right)'\right]^2\,,\\
&0=z_3''+\left(\ln\frac{c_1c_2^3}{c_3}\right)'\ z_3'-\frac{c_3^2}{c_2^2}\,,
\end{split}
\eqlabel{z2z3}
\end{equation}
should vanish at the boundary, \ie as $\rho\to 0$, and remain regular at the horizon,
\ie as $\rho\to \rho_H$. 

The on-shell regularized bulk action $S^\rho_{bulk}[\varphi]$ is a total
derivative\footnote{See \cite{Buchel:2004hw} for the $\caln=2^*$ gauge theory and \cite{Buchel:2011cc}
for the cascading gauge theory.},
however, we need to discard the contribution from the horizon \cite{Son:2002sd}.

The subtle piece in the boundary functional \eqref{pres4} comes from $S^{counter}[\varphi]$.
This boundary counterterm action includes finite counterterms, that can lead to renormalization
ambiguities in $\hg_\omega$ and $\hg_q$. Suppose that our holographic model
has operators of conformal dimension $\Delta=\{2,3\}$, the  dual bulk gravitational
scalars are correspondingly $\{\alpha,\chi\}$,
with coupling constants $\lambda_2$ and $\lambda_3$. Thus,
close to the $AdS_5$ boundary we
have\footnote{In the $\caln=2^*$ model $\lambda_2\sim m_b^2$ and $\lambda_3\sim m_f$.}
\begin{equation}
\alpha=\lambda_2\ \rho^2\ln\rho +\calo(\rho^2)\,,\qquad \chi=\lambda_3\ \rho+\calo(\rho^3\ln\rho)\,.
\eqlabel{uvexpansion}
\end{equation}
Holographic models with such operator content have finite counterterms such as 
\cite{Buchel:2012gw}
\begin{equation}
S^{counter}_{finite}=\frac{1}{16\pi G_5}\int_{\del\calm_5} dx^4\ \sqrt{-\gamma}\ R_4^\gamma \biggl(
\delta_b\ \frac{\alpha}{\ln \rho}+\delta_f\ \chi^2
\biggr)\,,
\eqlabel{fini1}
\end{equation}
where $\gamma_{ij}(\rho)$ is a four dimensional metric on the regularized boundary
$\del\calm_5$, and $R^\gamma_4$ is the Ricci scalar constructed from this metric, with $\rho$
being treated as an external parameter. $\delta_b$ and $\delta_f$ are arbitrary constants
specifying the renormalization scheme. Evaluation of $S^{counter}_{finite}$ on the bulk fluctuation
$\varphi$ produces\footnote{We set the asymptotic $AdS_5$ radius $L=1$.} nonvanishing as $\rho\to 0$ terms coming from $R^\gamma_4[\varphi]$, 
\begin{equation}
\begin{split}
R^\gamma_4[\varphi]=&\frac{1}{2c_2^2}\varphi \del^2_{zz}\varphi -\frac{1}{2c_1^2}\varphi\del^2_{tt}\varphi
+\calo(\varphi^4)\bigg|_{\rho\to 0}\\
=&\frac{\rho^2}{2}\ \int\frac{d\omega dq}{(2\pi)^2}\ 
\varphi^b(-\omega,-q)\ [\omega^2-q^2]\ \varphi^b(\omega,q)\ +\ \calo((\varphi^b)^4)\,,
\end{split}
\eqlabel{fini2}
\end{equation}
resulting in the renormalization scheme dependence of the retarded Green's function \eqref{gr}
\begin{equation}
G_{R,finite}^{xy,xy}(\omega,q)=\frac{\delta_b  \lambda_2+\delta_f\lambda_3^2}{16\pi G_5}\
\biggl(q^2-\omega^2\biggr)\,.
\eqlabel{fini3}
\end{equation}
The $(\delta_b  \lambda_2+\delta_f\lambda_3^2)$ factor immediately implies that
the gravitational susceptibility $\kappa$ is renormalization scheme dependent, as in  
\eqref{schemek}. It is clear that evaluating the logarithmic derivative $\frac{d\kappa}{d\ln T}$
completely removes this scheme dependence. Furthermore, the $(q^2-\omega^2)$ structure
of the renormalization ambiguity in \eqref{fini3} implies that the sum $T^2 (\hg_\omega+\hg_q)$
from \eqref{gr} is always renormalization scheme independent. As a result, see \eqref{gsum},
the shear relaxation time $\tau_\pi$ is renormalization scheme unambiguous as
well\footnote{The shear viscosity $\eta$ is universal $\frac{\eta}{s}=\frac{1}{4\pi}$
\cite{Buchel:2003tz}, and
is renormalization scheme independent \cite{Buchel:2004qq}.}.

\section{Cascading QGP}
\label{cqgp}

\begin{figure}[t]
\begin{center}
\psfrag{x}[cc][][0.7][0]{$\left(\frac 13-c_s^2\right)$}
\psfrag{y}[bb][][0.7][0]{$4\pi^2 T\kappa/s$}
\psfrag{z}[tt][][0.7][0]{$4\pi^2 T\kappa/s$}
\psfrag{a}[cc][][0.7][0]{$\frac T\Lambda$}
\includegraphics[width=2.8in]{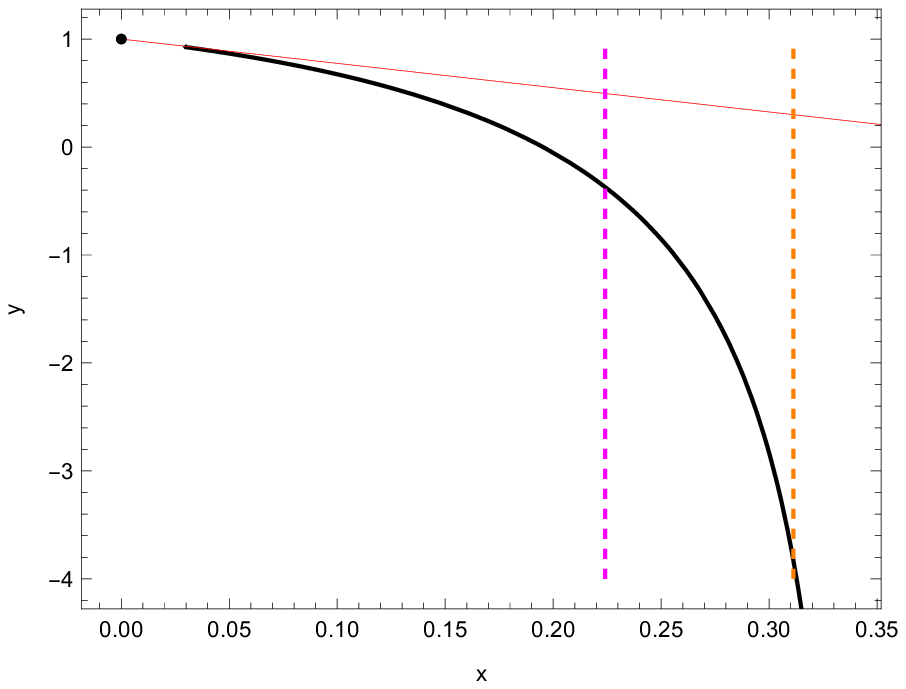}
\includegraphics[width=2.8in]{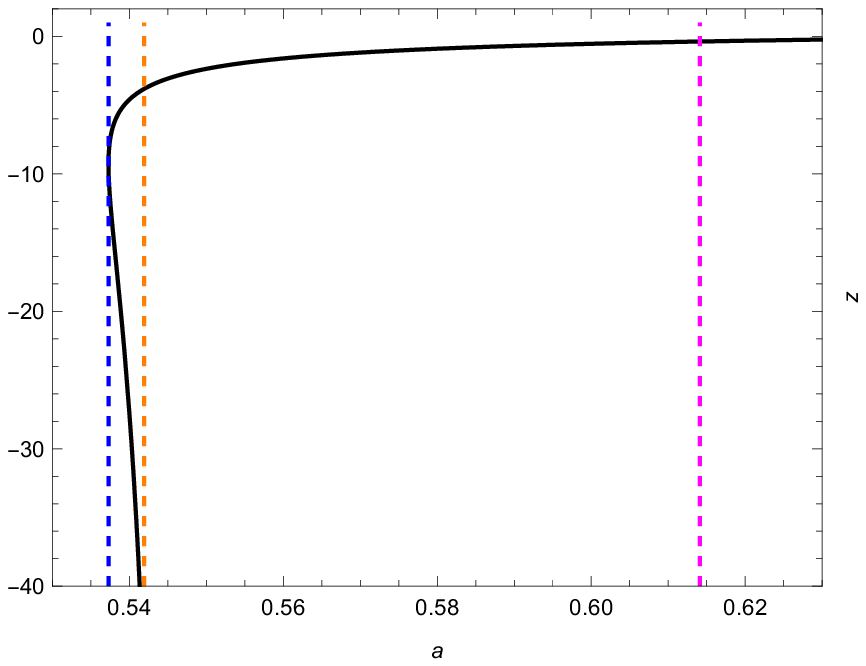}
\end{center}
  \caption{The normalized gravitational susceptibility $\kappa$ of the cascading QGP
  as a function of the nonconformal deformation parameter $\left(\frac 13-c_s^2\right)$ (the left panel),
  and the ratio $\frac T\Lambda$ (the right panel). The red line is the leading near-conformal approximation,
  and the vertical lines represent various phase transitions in this QGP.
  } \label{ktkappa}
\end{figure}

\begin{figure}[t]
\begin{center}
\psfrag{x}[cc][][0.7][0]{$\left(\frac 13-c_s^2\right)$}
\psfrag{y}[bb][][0.7][0]{$\calr_{\tau_\pi} $}
\psfrag{z}[tt][][0.7][0]{$\calr_{\tau_\pi} $}
\psfrag{a}[cc][][0.7][0]{$\frac T\Lambda$}
\includegraphics[width=2.8in]{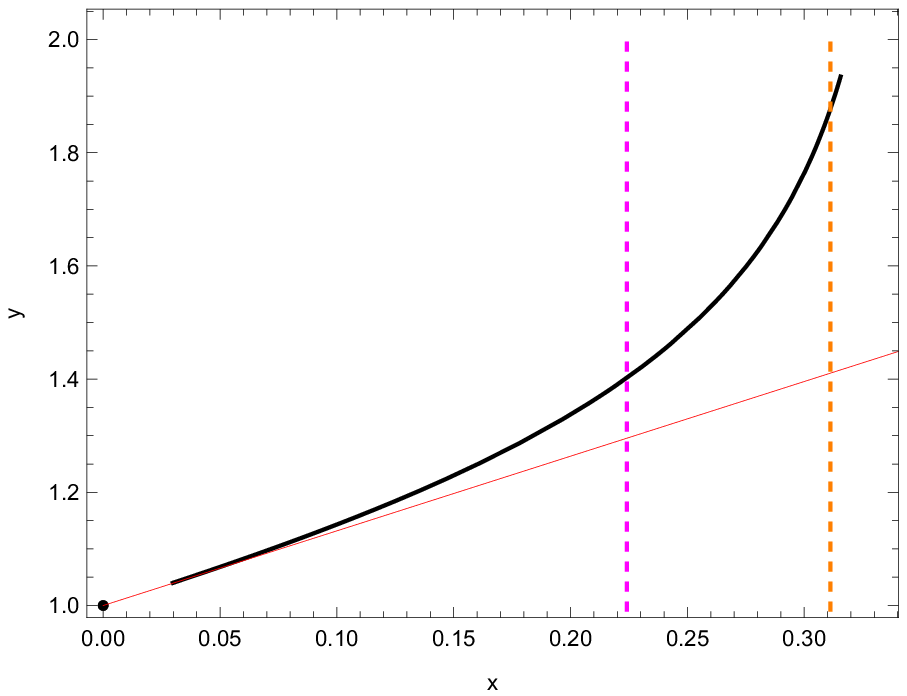}
\includegraphics[width=2.8in]{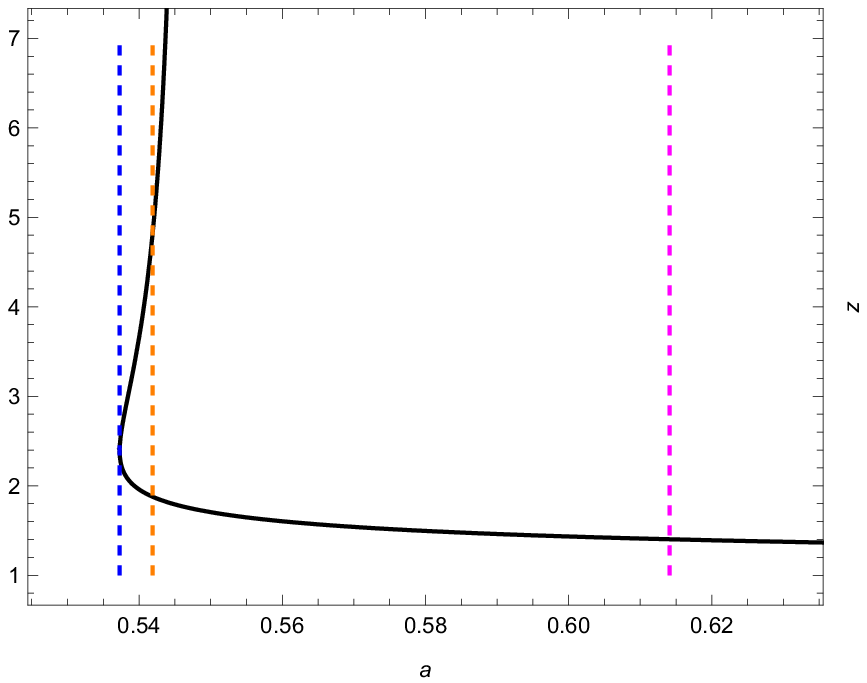}
\end{center}
  \caption{The normalized shear relaxation time  $\calr_{\tau_\pi}$ \eqref{defrtau} of the cascading QGP
  as a function of the nonconformal deformation parameter $\left(\frac 13-c_s^2\right)$ (the left panel),
  and the ratio $\frac T\Lambda$ (the right panel). The red line is the leading near-conformal approximation,
  and the vertical lines represent various phase transitions in this QGP.} \label{kttau}
\end{figure}

\begin{figure}[t]
\begin{center}
\psfrag{y}[bb][][0.7][0]{${\hat\kappa}/T^2$}
\psfrag{z}[tt][][0.7][0]{$T^{-1}{d\hat\kappa}/{dT} $}
\psfrag{x}[cc][][0.7][0]{$\frac T\Lambda$}
\includegraphics[width=2.8in]{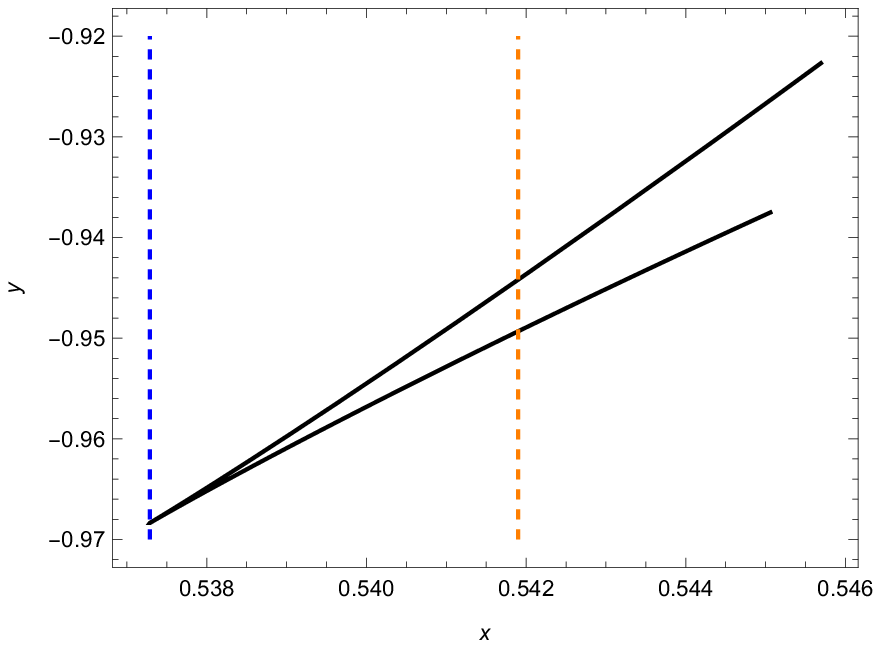}
\includegraphics[width=2.8in]{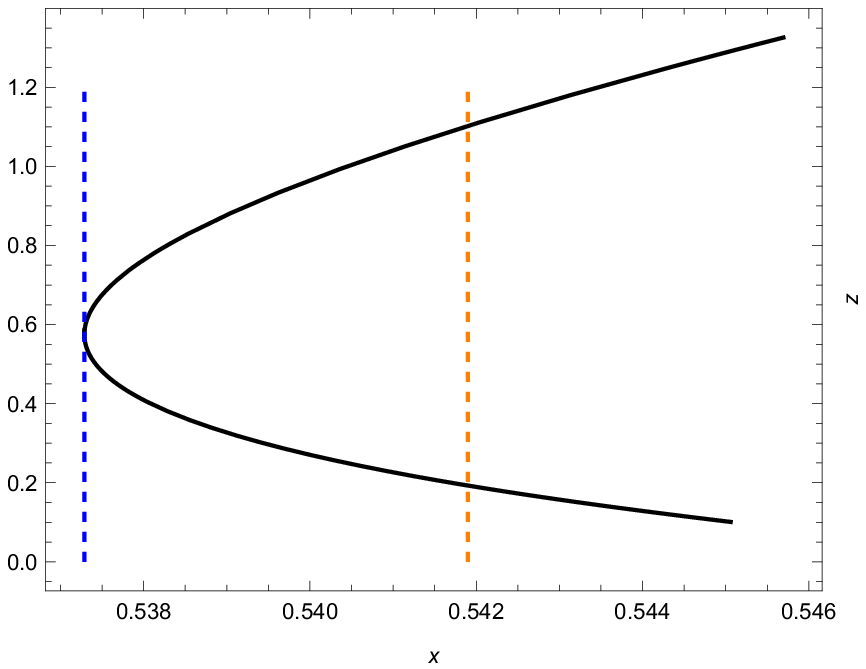}
\end{center}
  \caption{Behavior of $\hat\kappa\equiv 16\pi G_5 \kappa$ close
  to criticality, \ie as $T\to T_u+0$, see \eqref{casu}.
$T_u$ is represented by the vertical dashed blue line. 
The leading non-analytic term of
  the cascading gauge theory gravitational susceptibility 
  close to criticality is $\kappa_{non-analytic}\propto \pm (T-T_u)^{3/2}$.} \label{ktcrit}
\end{figure}

For a recent review of the cascading gauge theory see 
\cite{Buchel:2021yay}. In this section we will follow notations of the above
reference. We omit the technical details and highlight the results only.

The effective five-dimensional gravitational action used to describe
the chirally symmetric\footnote{The cascading QGP with spontaneously broken chiral symmetry is
unstable \cite{Buchel:2018bzp}.}
cascading gauge theory plasma contains the Einstein-Hilbert term, and four scalars dual to operators
of conformal dimensions $\Delta=\{4,4,6,8\}$. There are no relevant operators, and thus,
while the holographic renormalization of the theory is not unique \cite{Aharony:2005zr},
the gravitational susceptibility $\kappa$ of the theory is renormalization scheme independent.

The black brane dual to the chirally symmetric phase of the cascading QGP is characterized 
by 13 parameters (see eqs.~(A.55) and (A.58) of  \cite{Buchel:2021yay}):
\begin{equation}
\begin{split}
&{\rm UV:}\qquad \{K_0\,,\, f_{a,1,0}\,,\, f_{4,0}\,,\, f_{c,4,0}\,,\, g_{4,0}\,,\, f_{a,6,0}\,,\,
f_{c,8,0}\}\,,\\
&{\rm IR:}\qquad \{f_{a,0}^h\,,\, f_{c,0}^h\,,\, h_0^h\,,\, K_{1,0}^h\,,\, g_0^h\,,\, f_1^h\}\,,
\end{split}
\eqlabel{ksuv}
\end{equation}
where $K_0$ sets the strong coupling scale of the cascading gauge theory\footnote{We work
in the computation scheme with $P=g_s=1$.}  (see eq.~(2.48) of  \cite{Buchel:2021yay}):
\begin{equation}
\Lambda^2=\sqrt{2}\ e^{-K_0}\,.
\eqlabel{lambdacas}
\end{equation}
Parameters \eqref{ksuv} determine the thermodynamics of the theory (see eqs.~(A.59), (A.86), (A.92) of
\cite{Buchel:2021yay}):
\begin{equation}
\begin{split}
&8\pi G_5\ \epsilon=-\frac 32 f_{4,0}+\frac 32 f_{c,4,0}\,,\qquad 8\pi G_5\ P=-\frac 12 f_{4,0}-\frac 32 f_{c,4,0}\,,
\\
&4 G_5\ s=(f_{a,0}^h)^2\sqrt{f_{c,0}^h h_0^h}\,,\qquad T=\frac{f_1^h}{4\pi \sqrt{h_0^h}}\,.
\end{split}
\eqlabel{thermoks}
\end{equation}
For a given black brane geometry, a solution of \eqref{z2z3} is further characterized by 4 parameters:
\begin{equation}
\begin{split}
&{\rm UV:}\qquad \{z_{2,4,0}\,,\, z_{3,4,0}\}\,,\\
&{\rm IR:}\qquad \{z_{2,0}^h\,,\, z_{3,0}^h\}\,.
\end{split}
\eqlabel{ksgr}
\end{equation}
Implementing the holographic framework of section \ref{framework}, we confirm the general
structure of the retarded Green's function \eqref{gr}, and identify
\begin{equation}
\begin{split}
&16\pi G_5\ \hg_w=\frac{\pi^2 h_0^h}{2(f_1^h)^2}\ \biggl(f_{a,1,0}^2(6 K_0-7)-128 z_{2,4,0}\biggr)\,,
\\
&16\pi G_5\ \hg_q=-\frac{\pi^2 h_0^h}{2(f_1^h)^2}\ \biggl(f_{a,1,0}^2(6 K_0-7)+128 z_{3,4,0}\biggr)\,.
\end{split}
\eqlabel{gammaks}
\end{equation}

In fig.~\ref{ktkappa} we present the cascading QGP normalized gravitational susceptibility
$\frac{4\pi^2T\kappa}{s}$
as a function of the {\it universal} non-conformal deformation\footnote{This parameter is useful
in comparing different holographic models among themselves, and with the lattice QCD data (when available).
Its use was originally advocated for in \cite{Buchel:2007mf}.} 
$\left(\frac 13 -c_s^2\right)$ (the left panel) and as a function 
of model-specific $\frac T\Lambda$. The black dot indicates holographic $\caln=4$ SYM result \eqref{ksym}.
The red line is the independently computed\footnote{The thermal state of the
cascading QGP is constructed perturbatively in the limit $\ln\frac{T}{\Lambda}\gg 1$, see appendix D of
\cite{Buchel:2021yay}, followed by the corresponding perturbative solution of \eqref{z2z3}.} leading perturbative near-conformal approximation.
It agrees with an accuracy of $\sim 10^{-6}$ with the analytic result of \cite{Bigazzi:2010ku},
\begin{equation}
\frac{4\pi^2 T\kappa}{s}=1-\frac 94\left(\frac 13 -c_s^2\right)+\calo\left((1-3c_s^2)^2\right)\,.
\eqlabel{big1}
\end{equation}
The vertical dashed magenta line
indicates the first-order confinement/deconfinement phase transition at $T=T_c$,
and the vertical dashed orange line indicates the second-order chiral symmetry breaking
phase transition at $T=T_\csb$. Finally, the vertical dashed blue line indicates the terminal
temperature of the chirally symmetric phase of the cascading gauge theory plasma, see \eqref{casu}.

In fig.~\ref{kttau} we present the results for the normalized shear relaxation time $\calr_{\tau_\pi}$
\eqref{defrtau} of the cascading gauge theory plasma. Here, the agreement with the leading near-conformal
analytic result of \cite{Bigazzi:2010ku} is $\sim 3\cdot 10^{-6}$,
\begin{equation}
\calr_{\tau_\pi}=1+\frac {9(16-\pi^2)}{32(2-\ln 2)}\left(\frac 13 -c_s^2\right)+\calo\left((1-3c_s^2)^2\right)\,.
\eqlabel{big2}
\end{equation}

In fig.~\ref{ktcrit} we focus on the behavior of the cascading gauge theory susceptibility
\begin{equation}
\hat\kappa\equiv 16\pi G_5\ \kappa  
\eqlabel{defhatkappa}
\end{equation}
close to criticality, see \eqref{casu}: the left panel presents dimensionless quantity
$\frac{\hat\kappa}{T^2}$, and the right panel shows its temperature derivative.
From the plots it is clear the that the near-critical susceptibility of the cascading
QGP is given by \eqref{ktkks}; given \eqref{gr},  the latter implies \eqref{castaucr}.

\section{$\caln=2^*$ QGP}
\label{n2qgp}

In this section we follow notations of \cite{Buchel:2007vy}.
We omit the technical details and highlight the results only.

The effective five-dimensional gravitational action used to describe
$\caln=2^*$ gauge theory plasma contains the Einstein-Hilbert term, and two scalars dual to operators 
of conformal dimensions $\Delta=\{2,3\}$. Following the general discussion in section \ref{framework},
we expect two-parameter family of the renormalization scheme dependence of its gravitational susceptibility. 

The black brane dual to $\caln=2^*$ QGP is characterized 
by 8 parameters (see eqs.~(2.19) and (2.31) of  \cite{Buchel:2007vy}):
\begin{equation}
\begin{split}
&{\rm UV:}\qquad \{\hat{\delta}_3\,,\, \rho_{11}\,,\, \rho_{10}\,,\, \chi_0\,,\, \chi_{10}\}\,,\\
&{\rm IR:}\qquad \{a_h\,,\, r_0\,,\, c_0\}\,,
\end{split}
\eqlabel{n2uv}
\end{equation}
where $\{\rho_{11},\chi_0\}$ are the mass parameters of $\caln=2^*$  gauge theory
(see eq.~(3.12) of  \cite{Buchel:2007vy}):
\begin{equation}
\rho_{11}=\frac{\sqrt{2}}{24\pi^2}\ e^{-6 a_h}\ \left(\frac{m_b}{T}\right)^2\,,\qquad
\chi_0=\frac{1}{2^{3/4}\pi}\ e^{-3a_h}\ \left(\frac{m_f}{T}\right)\,.
\eqlabel{mn2}
\end{equation}
Parameters \eqref{n2uv} determine the thermodynamics of the theory (see eqs.~(2.36)-(2.39) of
\cite{Buchel:2007vy}):
\begin{equation}
\begin{split}
16\pi G_5\ P=&\frac12 \hat\delta_3^4 \biggl(
1+\rho_{11}^2 \biggl(24 \ln 2-96 \ln\hat\delta_3+16 \delta_2+24\biggr)
+2 \chi_{10} \chi_0^2-24 \rho_{10} \rho_{11}\\& +\chi_0^4 \left(-\frac23 \ln2+\frac83 \ln\hat\delta_3+\delta_1+\frac{10}{9}\right)
\biggr)\,,\\
16\pi G_5\ \epsilon=&2\hat\delta_3^4-16\pi G_5\  P\,,\qquad 4 G_5\ s = \hat\delta_3^3e^{3a_h}\,,\qquad
T=\frac{\hat\delta_3}{2\pi} e^{-3a_h}\,.
\end{split}
\eqlabel{thermon2}
\end{equation}
In \eqref{thermon2} the arbitrary constants $\delta_1$ and $\delta_2$ introduce the scheme dependence
to one-point correlation function of the $\caln=2^*$ boundary stress-energy tensor. With Minkowski space
supersymmetry, \ie when $m_b^2=m_f^2$, and correspondingly $\chi_0^2=6\rho_{11}$, the supersymmetry
preserving renormalization requires
\begin{equation}
0=9\delta_1+6+4\delta_2\,.
\eqlabel{n2susy}
\end{equation}
For a given black brane geometry, a solution of \eqref{z2z3} is further characterized by 4 parameters:
\begin{equation}
\begin{split}
&{\rm UV:}\qquad \{z_{2,2,0}\,,\, z_{3,2,0}\}\,,\\
&{\rm IR:}\qquad \{z_{2,0}^h\,,\, z_{3,0}^h\}\,.
\end{split}
\eqlabel{n2gr}
\end{equation}
Implementing the holographic framework of section \ref{framework}, we confirm the general
structure of the retarded Green's function \eqref{gr}, and identify
\begin{equation}
\begin{split}
&16\pi G_5\ \hg_w=\pi^2 e^{6 a_h} \biggl(-\frac{2\sqrt{2}}{9}  \biggl(12 \ln\hat\delta_3+9 \delta_3+5-3 \ln2\biggr)
\chi_0^2-8 \sqrt{2} \delta_4 \rho_{11}-4 z_{2,2,0}
\biggr)\,,
\\
&16\pi G_5\ \hg_q=-\pi^2 e^{6 a_h} \biggl(
-\frac{2\sqrt{2}}{9} \biggl(12 \ln\hat\delta_3+9 \delta_3+5-3 \ln2\biggr) \chi_0^2-8 \sqrt{2} \delta_4 \rho_{11}+4 z_{3,2,0}\biggr)\,.
\end{split}
\eqlabel{gamman2}
\end{equation}
Arbitrary constants $\delta_3\leftrightarrow \delta_f$ and $\delta_4\leftrightarrow \delta_b$ introduce the
renormalization scheme dependence in accordance with the
general discussion in section \ref{framework}.

An interesting feature of the $\caln=2^*$ QGP with $m_b^2=m_f^2$ is that the limit
$\frac{T}{m_b}\to 0$ is given by the conformal thermodynamics of a certain five-dimensional
theory, compactified on $S^1$ \cite{HoyosBadajoz:2010td}. The local properties of plasma,
such as the transport coefficients, are unaffected by the compactification.
The precise matching of the $\caln=2^*$  thermodynamics in the $\frac{T}{m_b}\to 0$ limit
with that of the $CFT_5$ thermodynamics was explained in \cite{Buchel:2020vkv};
it can be easily extended to the matching of the Green's functions, with the
(perhaps the obvious) result:
\begin{equation}
\begin{split}
&\lim_{{T}/{m_b}\to 0} \calr_\kappa\bigg|_{\caln=2^*,m_b=m_f}\ = \calr_\kappa\bigg|_{CFT_5}\,,\\
&\lim_{{T}/{m_b}\to 0} \calr_{\tau_\pi}\bigg|_{\caln=2^*,m_b=m_f}\ = \calr_{\tau_\pi}\bigg|_{CFT_5}\,,
\end{split}
\eqlabel{matching}
\end{equation}
where in view of the renormalization scheme dependence of the gravitational
susceptibility of the $\caln=2^*$ QGP we use \eqref{defcalrk}.
The relaxation time $\tau_\pi$ of the $CFT_5$ plasma was computed in  \cite{Haack:2008cp}
\begin{equation}
\begin{split}
&\tau_\pi T\bigg|_{CFT_5}=\frac{5}{8\pi}\biggl(2-\frac \pi5
\sqrt{1-\frac{2}{\sqrt{5}}}+\frac{1}{\sqrt{5}}\ \coth^{-1}\sqrt{5}-\frac 12\ln 5
\biggr)\\
&\Longrightarrow\qquad \calr_{\tau_\pi}\bigg|_{CFT_5}=1.15(4)\,.
\end{split}
\eqlabel{tau5}
\end{equation}
We reproduce \eqref{tau5}, and additionally find
\begin{equation}
\calr_{\tau_\pi}\bigg|_{CFT_5}=\frac 54\,.
\eqlabel{k5}
\end{equation}

\begin{figure}[t]
\begin{center}
\psfrag{x}[cc][][0.7][0]{$\left(\frac 13-c_s^2\right)$}
\psfrag{y}[bb][][0.7][0]{$\calr_{\kappa} $}
\psfrag{z}[tt][][0.7][0]{$\calr_{\kappa} $}
\includegraphics[width=2.8in]{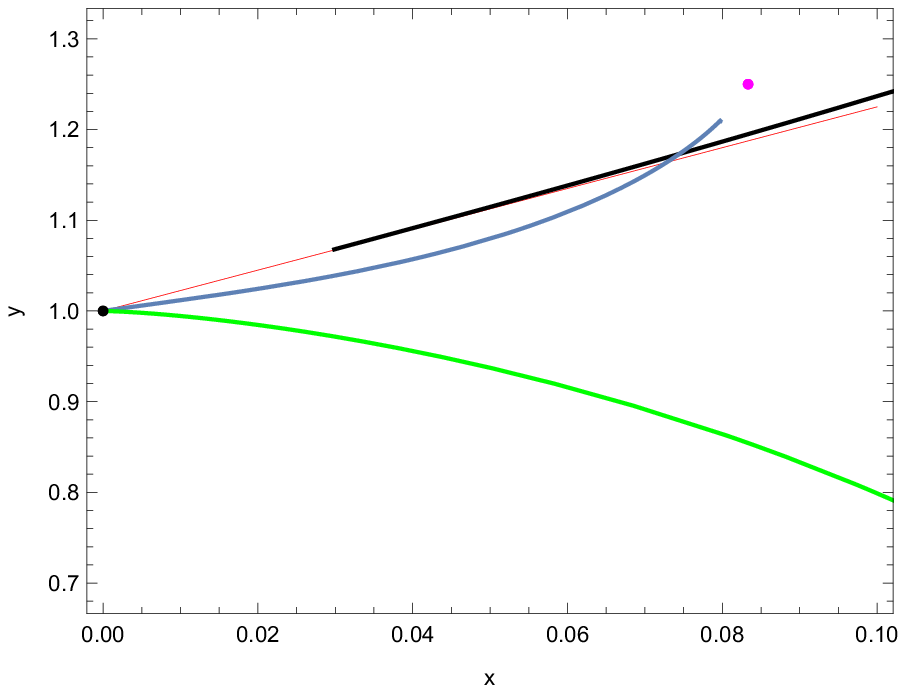}
\includegraphics[width=2.8in]{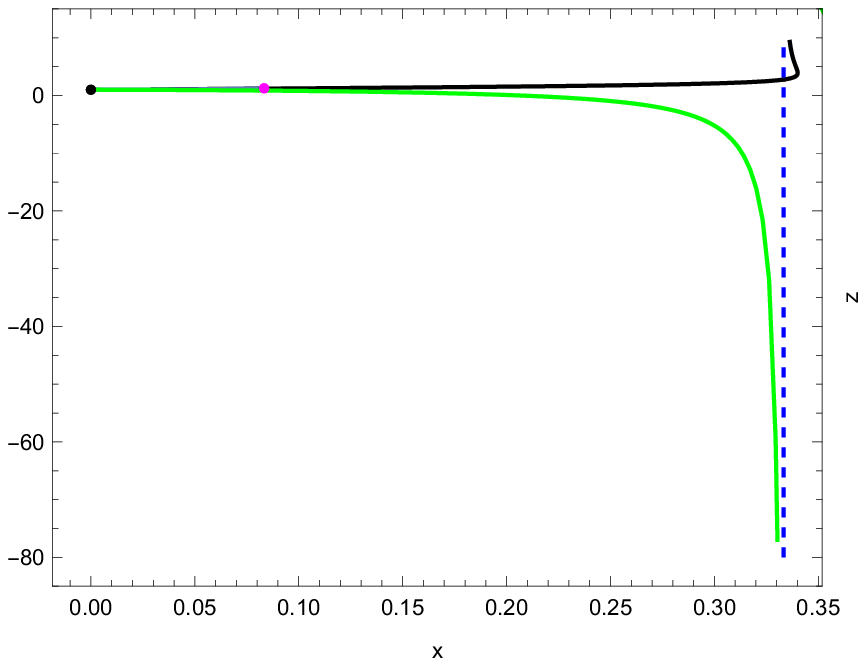}
\end{center}
  \caption{The gravitational susceptibility parameter $\calr_\kappa$, see \eqref{defcalrk},
  for the $\caln=2^*$ plasma with $m_b^2=m_f^2$ (the grey curve),
the $\caln=2^*$ plasma with $m_b\ne 0$ and $m_f=0$ (the green curve),
and the cascading QGP (the black curve). The vertical dashed blue line identifies the
critical behavior as $c_s^2\to 0$. Notice that while  $\calr_\kappa$
diverges for the $\caln=2^*$ QGP with $m_f=0$,
it remains finite for the cascading gauge theory plasma.} \label{n2kappa}
\end{figure}

\begin{figure}[t]
\begin{center}
\psfrag{x}[cc][][0.7][0]{$c_s^2$}
\psfrag{y}[bb][][0.7][0]{$1/\calr_{\kappa} $}
\psfrag{z}[tt][][0.7][0]{$m_b^2/T^2$}
\includegraphics[width=2.8in]{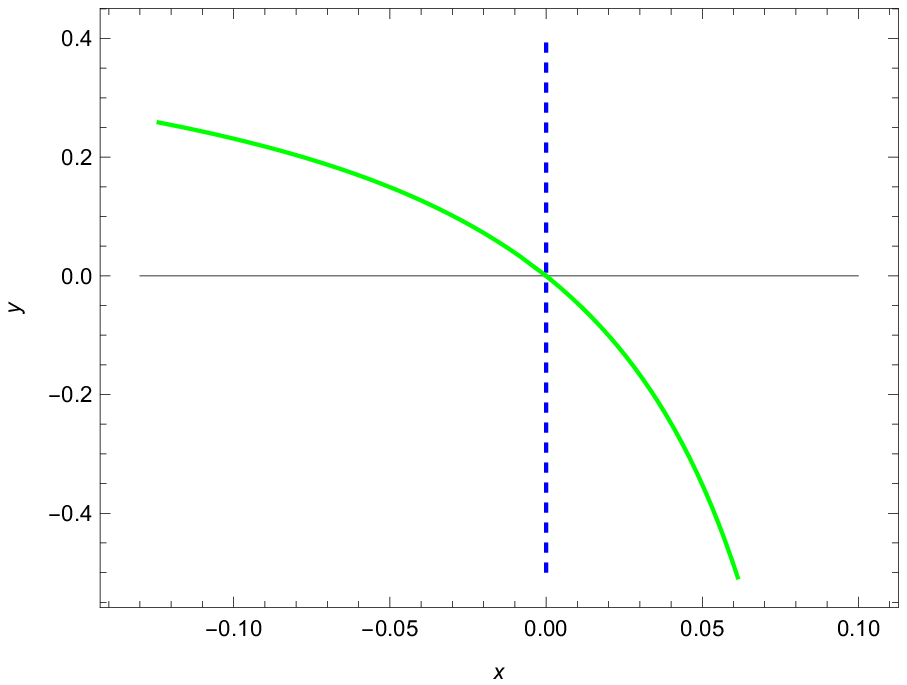}
\includegraphics[width=2.8in]{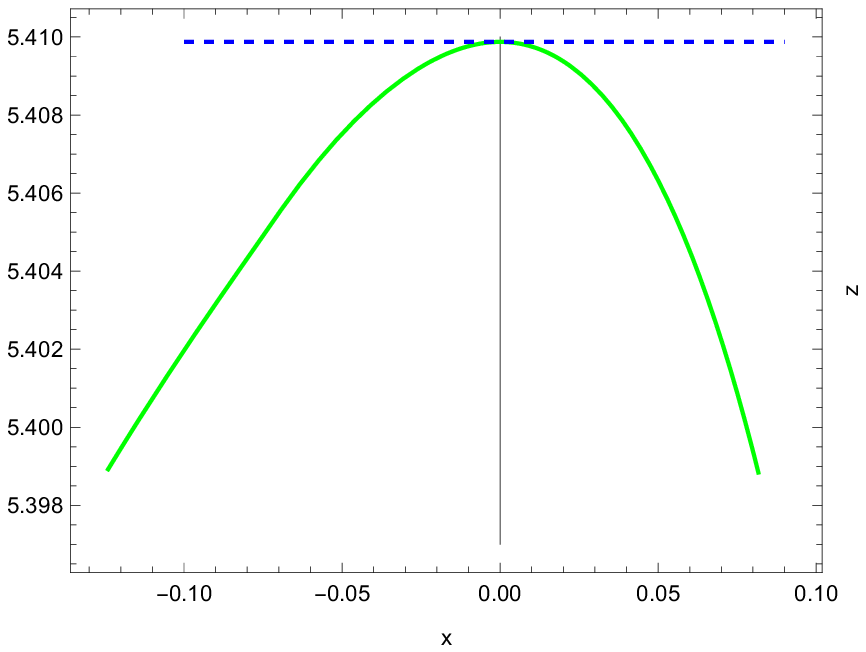}
\end{center}
  \caption{The $\caln=2^*$ QGP with $m_f=0$ has a terminal temperature $T_{crit}$ \eqref{n2crit},
  represented by the horizontal dashed blue line. Close to criticality, the gravitational
  susceptibility parameter $\calr_\kappa$ of the model 
  diverges as $\calr_\kappa\propto -\frac{1}{c_{s}^2}\propto \mp \frac{1}{\sqrt{T-T_{crit}}}$.} \label{n2critk}
\end{figure}

\begin{figure}[t]
\begin{center}
\psfrag{x}[cc][][0.7][0]{$(\frac 13-c_s^2)$}
\psfrag{a}[cc][][0.7][0]{$c_s^2$}
\psfrag{y}[bb][][0.7][0]{$\calr_{\tau_\pi} $}
\psfrag{z}[tt][][0.7][0]{$1/\calr_{\tau_\pi}$}
\includegraphics[width=2.8in]{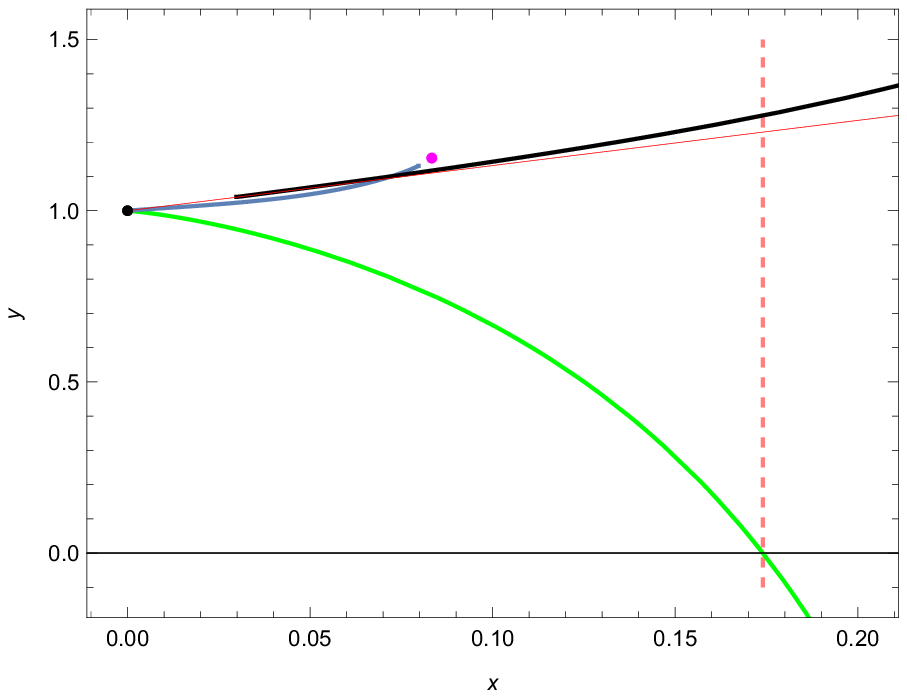}
\includegraphics[width=2.8in]{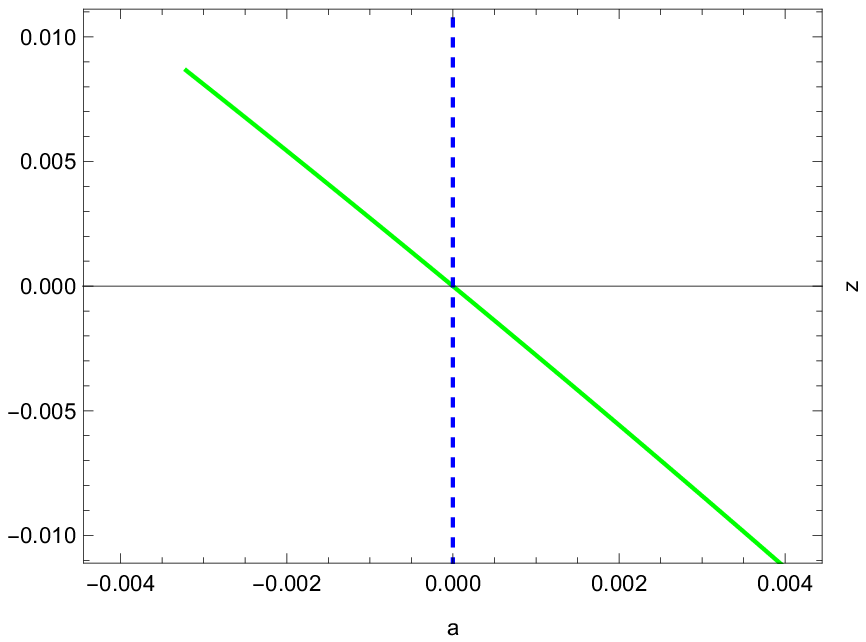}
\end{center}
  \caption{The left panel: the shear relaxation time parameter $\calr_{\tau_\pi}$, see \eqref{defrtau},  for all the models we study:
the $\caln=2^*$ plasma with $m_b^2=m_f^2$ (the grey curve),
the $\caln=2^*$ plasma with $m_b\ne 0$ and $m_f=0$ (the green curve),
and the cascading QGP (the black curve).
Notice that below some temperature (the vertical pink dashed line),
the shear relaxation time of the $\caln=2^*$ QGP with $m_f=0$ becomes negative. 
The right panel: the shear relaxation time
of the $\caln=2^*$ QGP with $m_f=0$ diverges as in \eqref{taucritn2} as $c_s^2\to 0$,
represented by the vertical dashed blue line.} \label{n2crittau}
\end{figure}

In fig.~\ref{n2kappa} we collect the gravitational susceptibility parameter
$\calr_{\kappa}$, see \eqref{defcalrk}, for all the models we study:
the $\caln=2^*$ plasma with $m_b^2=m_f^2$ (the grey curve),
the $\caln=2^*$ plasma with $m_b\ne 0$ and $m_f=0$ (the green curve),
and the cascading QGP (the black curve). The black dot indicates
the $\caln=4$ SYM result \eqref{ksym}, and the magenta dot indicates the
$CFT_5$ result \eqref{k5}. The red line is the near-conformal approximation
to $\calr_\kappa$ for the cascading gauge theory plasma. The vertical
dashed blue line identifies the critical behavior with the vanishing
speed of the sound waves, \ie $T\to T_{crit}$ \eqref{n2crit} for $\caln=2^*$ plasma
with $m_f=0$, and $T\to T_u$ \eqref{casu} for the cascading gauge theory plasma.

In fig.~\ref{n2critk} we highlight the critical behavior of $\calr_\kappa$ in
the $\caln=2^*$ QGP with $m_f=0$. Since in the critical region, see the right panel,
\begin{equation}
T-T_{crit}\ \propto\  (c_s^2)^2 \,,
\eqlabel{tcrits}
\end{equation}
and, see the left panel,
\begin{equation}
\calr_\kappa\ \propto\ -\frac{1}{c_s^2}\,, 
\eqlabel{tcrits6}
\end{equation}
we extract the divergent critical behavior of $\calr_\kappa$ as in \eqref{rcrit}. 

In fig.~\ref{n2crittau} we collect the shear relaxation time parameter
$\calr_{\tau_\pi}$, see \eqref{defrtau},  for all the models we study (the left panel):
the $\caln=2^*$ plasma with $m_b^2=m_f^2$ (the grey curve),
the $\caln=2^*$ plasma with $m_b\ne 0$ and $m_f=0$ (the green curve),
and the cascading QGP (the black curve). The black dot indicates
the $\caln=4$ SYM result \eqref{taupin4}, and the magenta dot indicates the
$CFT_5$ result \eqref{tau5}. The red line is the near-conformal approximation
to $\calr_{\tau_\pi}$ for the cascading gauge theory plasma.
Notice that the shear relaxation time of the $\caln=2^*$ plasma with $m_f=0$ becomes
negative for $T< 1.003(0) T_{crit}$, represented by the vertical dashed pink line.
In the right panel we show the critical behavior of the shear relaxation time of
the $\caln=2^*$ QGP with $m_f=0$ as $T\to T_{crit}$,
$\calr_{\tau_\pi}\propto -\frac{1}{c_s^2}$, leading to \eqref{taucritn2}.

\section*{Acknowledgments}
I would like to thank Francesco Bigazzi for pointing out \cite{Bigazzi:2010ku}.
Research at Perimeter Institute is supported in part by the Government
of Canada through the Department of Innovation, Science and Economic
Development Canada and by the Province of Ontario through the Ministry
of Colleges and Universities. This work is further supported by a
Discovery Grant from the Natural Sciences and Engineering Research
Council of Canada.

%\appendix

\bibliographystyle{JHEP}
\bibliography{kappa}

\end{document}